\documentstyle[epsfig]{mn}

\def\lesssim{\mathrel{\hbox{\rlap{\hbox{\lower4pt\hbox{$\sim$}}}\hbox{$<$}}}}
\def\gtrsim{\mathrel{\hbox{\rlap{\hbox{\lower4pt\hbox{$\sim$}}}\hbox{$>$}}}}

\title[Gamma-ray absorption during cosmic reionization]
{Probing intergalactic radiation fields during cosmic reionization through gamma-ray absorption
\thanks{Numerical data of the model results will be available at http://www-tap.scphys.kyoto-u.ac.jp/\~{ }inoue/hizabs/}
}
\author[S. Inoue, R. Salvaterra, T. R. Choudhury, A. Ferrara, B. Ciardi \& R. Schneider]
{Susumu Inoue$^{1}$
\thanks{E-mail: inoue@tap.scphys.kyoto-u.ac.jp},
Ruben Salvaterra$^2$, Tirthankar Roy Choudhury$^3$,
\newauthor
Andrea Ferrara$^{4}$, Benedetta Ciardi$^5$ and Raffaella Schneider$^6$\\
$^1$Department of Physics, Kyoto University, Oiwake-cho, Kitashirakawa, Sakyo-ku, Kyoto 606-8502, Japan\\
$^2$INAF, Osservatorio Astronomico di Brera, via E. Bianchi 46, 23807 Merate, Italy\\
$^3$Harish-Chandra Research Institute, Chhatnag Road, Jhunsi, Allahabad 211 019, India\\
$^4$Scuola Normale Superiore, Piazza dei Cavalieri 7, 56126 Pisa, Italy\\
$^5$Max-Planck-Institut f\"ur Astrophysik, K.-Schwarzschild-Str. 1, 85748 Garching, Germany\\
$^6$Osservatorio Astrofisico di Arcetri, Largo Enrico Fermi 5, 50125 Firenze, Italy}

\begin{document}

\date{Accepted ?. Received ?; in original form ?}

\pagerange{\pageref{firstpage}--\pageref{lastpage}} \pubyear{2009}

\maketitle

\label{firstpage}

\begin{abstract}
We discuss expectations for the absorption of high-energy gamma-rays by $\gamma\gamma$ pair production
with intergalactic radiation fields (IRFs) at very high redshifts ($z \sim 5-20$),
and the prospects thereof for probing the cosmic reionization era.
For the evolving IRF, a semi-analytical model
incorporating both Population II and Population III stars is employed,
which is consistent with a wide variety of existing high-$z$ observations
including QSO spectral measurements, {\it WMAP} Thomson depth constraints,
near-IR source count limits, etc.
We find that the UV IRF below the Lyman edge energy
with intensities in the range of a few times $10^{-19} {\rm \ erg \ cm^{-2} s^{-1} Hz^{-1} sr^{-1}}$
can cause appreciable attenuation above $\sim$12 GeV at $z \sim 5$, down to $\sim 6-8$ GeV at $z \gtrsim 8-10$.
This may be observable in the spectra of
blazars or gamma-ray bursts by the {\it Fermi} Gamma-ray Space Telescope,
or next generation facilities such as the Cherenkov Telescope Array, Advanced Gamma-ray Imaging System or 5@5,
providing invaluable insight into early star formation and cosmic reionization.
\end{abstract}


\section{Introduction}
\label{sec:intro}
Some time after the epoch of cosmic recombination at redshift $z \sim 1100$,
the bulk of the intergalactic gas in the universe must have been somehow reionized by $z \sim 6$,
as indicated observationally from the spectra of high-$z$ QSOs
and the polarization of the cosmic microwave background (CMB).
However, the sources, history and nature of this cosmic reionization process are still largely unknown,
as most of this redshift range has yet to be explored through direct observations.
Because the first stars and galaxies in the universe must have formed during this period,
the primary suspect is photoionization by UV radiation from such objects,
potentially involving metal-free, Population (Pop) III stars.
Alternative possibilities include mini-quasars, supernova remnants and dark matter decay.
Besides providing us with clues to such processes in the early universe,
cosmic reionization also profoundly affects the ensuing formation of stars and galaxies,
so elucidating this era is one of the most pressing issues in cosmology today
(see Barkana \& Loeb 2001; Ciardi \& Ferrara 2005; Fan et al. 2006, Choudhury 2009 for reviews).

In the majority of scenarios for reionization of hydrogen in the intergalactic medium (IGM),
the main protagonists are UV photons with energies above the Lyman edge ($\epsilon \ge \epsilon_{\rm LE}=13.6$ eV).
Although those with lower energies do not contribute to photoionization,
they are also crucial since
i) they give indications as to the strength and nature of the ionizing radiation,
ii) those in the Lyman-Werner band ($\epsilon=11.2-13.6$ eV)
can photodissociate $\rm H_2$ molecules and suppress early star formation (e.g. Ciardi \& Ferrara 2005), and
iii) Ly$\alpha$ photons ($\epsilon=10.2$ eV) can strongly affect the HI spin temperature
and the associated cosmological 21 cm signatures (e.g. Furlanetto et al. 2006).
Thus, having some observational means to probe the evolution of UV intergalactic radiation fields (IRFs)
\footnote{Although often referred to as ``extragalactic background light'' for lower $z$,
here we avoid the term ``background'', since IRFs
can be highly inhomogeneous in the reionization era, especially for $\epsilon \ge \epsilon_{\rm LE}$,
even though it turns out to be more or less uniform for the spectral regime relevant to $\gamma\gamma$ absorption; see below.}
in the cosmic reionization era would be of paramount importance,
complementing existing observations that probe the neutral or ionized gaseous components of the IGM.
However, direct detection of this diffuse emission from very high $z$ is extremely difficult if not impossible. 
\footnote{Earlier indications of a large contribution from Pop III stars to the local near-IR background
are now disfavored from J-band source count limits (Salvaterra \& Ferrara 2006).}

An indirect but powerful means of probing diffuse radiation fields
is through photon-photon ($\gamma\gamma$) absorption of high-energy gamma-rays
(e.g. Gould \& Schreder 1967; Stecker et al. 1992).
Gamma-rays with energy $E$ emitted from extragalactic sources
will be absorbed during intergalactic propagation by interacting with photons of the diffuse radiation field with energy $\epsilon$
to produce electron-positron pairs ($\gamma+\gamma \rightarrow e^+ + e^-$),
as long as there is sufficient opacity for energies satisfying the threshold condition $E \epsilon (1-\cos \theta) \ge 2 m_e^2 c^4$,
where $\theta$ is the incidence angle of the two photons.
The observed spectra of the gamma-ray sources should then exbihit corresponding attenuation features,
from which one can effectively infer or limit the properties of the diffuse radiation.
This method has been utilized in recent TeV observations of blazars by ground-based Cherenkov telescopes
to set important constraints on the extragalactic background light in the near infrared to optical bands
at relatively low $z$ (Aharonian et al. 2006; Albert et al. 2008).

As first discussed by Oh (2001; see also Rhoads 2001), UV IRFs
with sufficient intensities to cause IGM reionization are also likely to induce
significant $\gamma\gamma$ absorption in gamma-ray sources at $z \gtrsim 6$
at observed energies in the range of a few to tens of GeV.
However, these estimates
i) were made before {\it WMAP} observations indicating an early start of reionization and were limited to $z \le 10$,
and ii) did not include the possibility of metal-free Pop III stars,
which may have been active during the first epochs of star formation
and are more prodigious UV emitters compared to normal stars
\footnote{Note that $\gamma\gamma$ absorption measurements in low-$z$ blazars
have set strong constraints against a large contribution from Pop III stars to the local near-IR background
(Aharonian et al. 2006; see also Raue et al. 2009).}.
The recent launch of the {\it Fermi} satellite \footnote{http://fermi.gsfc.nasa.gov}
with the Large Area Telescope (LAT) operating in the $\sim 0.1-100$ GeV domain
motivates us to reevaluate the $\gamma\gamma$ absorption opacity at very high $z$,
incorporating more recent observational and theoretical developments concerning the cosmic reionization era.

For this purpose, we employ updated versions of the semi-analytical models of Choudhury \& Ferrara (2005; 2006),
which self-consistently describe inhomogeneous reionization of the IGM,
accounting for both Pop II and Pop III stars and their radiative and chemical feedback effects.
With only a few free parameters, they are able to fit a wide variety of high-$z$ observational data.
Using the evolving IRFs as predicted by these models,
the $\gamma\gamma$ opacity is evaluated for the redshift range $z=5-20$.
We also briefly assess the detectability of the resultant absorption features
in high-$z$ sources such as blazars or gamma-ray bursts (GRBs) with current and future gamma-ray facilities,
and the consequent implications.

\section{Intergalactic Radiation Field Model}
\label{sec:model}
The salient features of our semi-analytical models are as follows
(see Choudhury \& Ferrara 2005; 2006; 2007; Choudhury et al. 2008; Choudhury 2009 for more details):
(1) Adopting a lognormal distribution of IGM inhomogeneities (Miralda-Escud\'e et al. 2000),
the ionization and thermal histories of the neutral, HII and HeII phases of the IGM
are tracked simulaneously and self-consistently.
(2) The formation and evolution of dark matter halos are described by a Press-Schechter-based approach.
(3) Three types of radiation sources are considered:
a) metal-free Pop III stars with a Salpeter initial mass function (IMF) in the mass range $1-100 M_\odot$,
with spectra according to Schaerer (2002) and including nebular and Ly$\alpha$ emission lines (see Salvaterra \& Ferrara 2003);
b) low-metallicity ($Z=0.02 Z_\odot$) Pop II stars with spectra according to Bruzual \& Charlot (2003),
otherwise being the same as Pop III; and
c) QSOs with power-law spectra
and emissivity based on the observed luminosity function at $z<6$,
considering only those above the break luminosity (Choudhury et al. 2008).
(4) Pop II and Pop III stars each form from gas in virialized halos
with efficiencies $\varepsilon_{\rm *,II}$ and $\varepsilon_{\rm *,III}$, respectively,
and the corresponding escape fractions of ionizing photons from the host halos are 
parameterized by $f_{\rm esc,II}$ and $f_{\rm esc,III}$.
Included self-consistently are the consequent effects of
radiative feedback that suppresses star formation in sufficiently small halos,
as well as a ``genetic'' merger-tree-based treatment of chemical feedback
that induces the transition from Pop III to Pop II star formation (Schneider et al. 2006).

The free parameters of the model are $\varepsilon_{\rm *,II}$, $\varepsilon_{\rm *,III}$,
$\eta_{\rm esc}$, which fixes both $f_{\rm esc,II}$ and $f_{\rm esc,III}$,
and $\lambda_0$, related to the mean free path of ionizing photons due to HI in high-density regions.
\footnote{The adopted cosmological parameters are
$h=0.73$, $\Omega_m=0.24$, $\Omega_\Lambda=0.76$, $\Omega_b h^2=0.022$,
$\sigma_8=0.74$, $n_s=0.95$ and $dn_s/d\ln k=0$ (Spergel et al. 2007).}
These are ascertained so as to simultaneously reproduce a large set of high-$z$ observational data:
i) redshift evolution of Lyman-limit absorption systems;
ii) effective optical depths of the IGM for Ly$\alpha$ and Ly$\beta$ from QSO spectra;
iii) electron scattering optical depth $\tau_e$ from {\it WMAP} 3rd year results (Spergel et al. 2007);
\footnote{The fiducial model gives $\tau_e=0.07$, consistent with the 5th year results as well (Dunkley et al. 2009).}
iv) temperature of the mean IGM;
v) cosmic star formation history; and
vi) limits on J-band source counts from NICMOS HUDF.
In the fiducial, best-fit model,
\footnote{The relevant parameters are
$\varepsilon_{\rm *,II}=0.1$, $\varepsilon_{\rm *,III}=0.02$, $f_{\rm esc,II}=0.0578$ and $f_{\rm esc,III}=0.54$.}
H reionization begins rapidly at $z \sim 15$, initially driven by Pop III stars,
and is 90 \% complete by $z \sim 8$.
Thereafter it is slowed down by feedback effects
and taken over by Pop II stars at $z \sim 7$,
finally reaching completion by $z \sim 6$ (see Fig.2 of Choudhury 2009).
The cosmic star formation rate is always dominated by Pop II stars
and is at the level of $0.05-0.08 \ {\rm M_\odot \ yr^{-1} Mpc^{-3}}$
for $z \sim 6-8$ (Fig.2(b) of Choudhury 2009),
in line with that deduced from observed GRB rates (e.g. Salvaterra et al. 2008, Kistler et al. 2009).

\begin{figure}
\centering
\epsfig{file=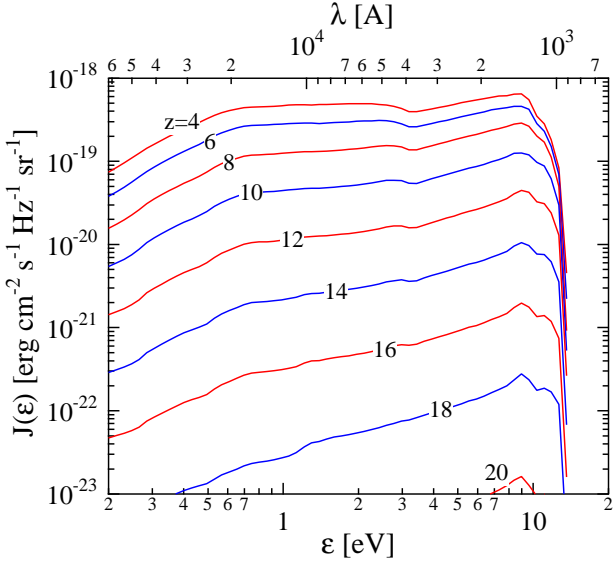,width=0.50\textwidth}
\caption{
Volume-averaged intensity of the intergalactic radiation field $J(\epsilon)$
vs. energy $\epsilon$ (or wavelength $\lambda$)
at redshifts $z$ as labelled for the fiducial model.
}
\label{fig:bg}
\end{figure}

Shown in Fig.\ref{fig:bg} is the volume-averaged intensity of the IRF as calculated from this model,
which declines monotonically with $z$ following the evolution of the star formation rate (SFR).
We caution that at $z \gtrsim 6$ before intergalactic HII regions have completely overlapped,
the IRF is expected to be inhomogeneous and fluctuating along different lines of sight,
particularly strongly for $\epsilon \ge \epsilon_{\rm LE}$.
However, it is also evident that ionizing photons are strongly absorbed by the neutral IGM
and the mean IRF spectrum cuts off very sharply above $\epsilon_{\rm LE}$,
so that this portion has negligible effects on the $\gamma\gamma$ opacity (Madau \& Phinney 1996, Oh 2001, Rhoads 2001).
On the other hand, UV radiation with $\epsilon < \epsilon_{\rm LE}$
have much longer mean free paths in the IGM,
and the notion of a nearly uniform and isotropic background may still be appropriate
for this regime, which is also the most relevant for $\gamma\gamma$ absorption.
Of particular note is the band $\epsilon = 10.2-13.6$ eV
where the spectrum dips somewhat due to blanketing by the Lyman series lines,
but which should nevertheless be very important for the $\gamma\gamma$ opacity.

Being optimized for the cosmic reionization era,
the main shortcoming of the present model is that
it does not account for Pop I stars or dust that can become important at lower $z$.
Our IRF calculations are available only for $z \ge 4$,
and may be somewhat less reliable near $z \sim 4$
as the comparison with observations has not been as thorough as for $z > 5$. 
A more complete model describing the evolution of the IRF at all redshifts awaits future studies.

\section{Gamma-ray Absorption Opacity}
\label{sec:pairabs}
\begin{figure}
\centering
\epsfig{file=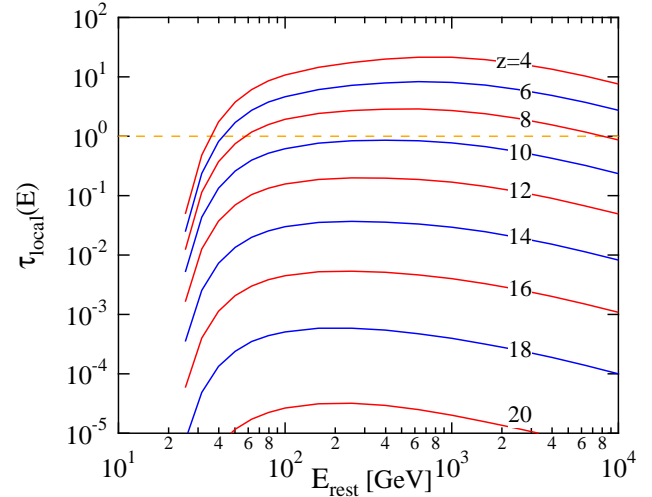,width=0.50\textwidth}
\caption{
Local $\gamma\gamma$ optical depth $\tau_{\rm local}(E)$ vs. rest-frame gamma-ray energy $E_{\rm rest}$
at redshifts $z$ as labelled for the fiducial model.
The contribution from the CMB is not shown.
}
\label{fig:tauloc}
\end{figure}

We first estimate the ``local $\gamma\gamma$ optical depth'' at each $z$
by the optical depth across a Hubble radius $l_H(z)=c/H_0 (\Omega_m (1+z)^3+\Omega_\Lambda)^{1/2}$,

\begin{eqnarray}
\tau_{\rm local} (z,  E) =  l_H(z) \int_{\epsilon_{\rm th}}^{\infty} d\epsilon \ n (\epsilon, z) \nonumber \\
                                            \times {1 \over 2} \int_{-1}^{1} d\mu (1 - \mu) \sigma_{\gamma\gamma} (E, \epsilon, \mu),                                 
\label{tauloc}
\end{eqnarray}
where $n(\epsilon, z)$ is the IRF photon number density per energy interval, $\mu=\cos \theta$,
$\epsilon_{\rm th} = 2 m_e^2 c^4/E (1-\mu)$ is the threshold energy,
and $\sigma_{\gamma\gamma} (E, \epsilon, \mu)$
is the $\gamma\gamma$ pair production cross section.
For given $E$, $\sigma_{\gamma\gamma}$ rises sharply from $\epsilon=\epsilon_{\rm th}$,
peaks at $\epsilon=2\epsilon_{\rm th}$, and then falls off as $\epsilon^{-1}$.
Thus $\tau_{\rm local}$ roughly mirrors the IRF spectrum at each $z$,
although its detailed features are smeared out.
Displayed in Fig.\ref{fig:tauloc} in terms of the rest-frame gamma-ray energy $E_{\rm rest}$,
we see that the opacity may be significant out to $z \sim 10$ for $E_{\rm rest} \sim 10^2-10^4$ GeV.
\footnote{
The contribution from the CMB (e.g. Stecker et al. 2006)
also becomes important at $E_{\rm rest} \gtrsim 2$ TeV,
but is irrelevant for our results below and not plotted in Fig.\ref{fig:tauloc}.}
Note the steep drop in $\tau_{\rm local}$ at $E_{\rm rest}<E_{\rm LE} \sim (m_e c^2)^2/\epsilon_{\rm LE} \simeq$ 18 GeV,
corresponding to the sharp cutoff in the IRF spectrum above the Lyman edge.
As pointed out by Oh (2001; see also Rhoads 2001),
this is crucial in that it allows appreciable contributions to the total $\gamma\gamma$ opacity
from higher $z$ even when the IRF intensity is relatively weaker,
and which should be uncontaminated from absorption at lower $z$.
However, we also see that due to the declining IRF intensity
together with the reduced path length,
the opacity from $z \gtrsim 10$ is likely to be quite small.

\begin{figure}
\centering
\epsfig{file=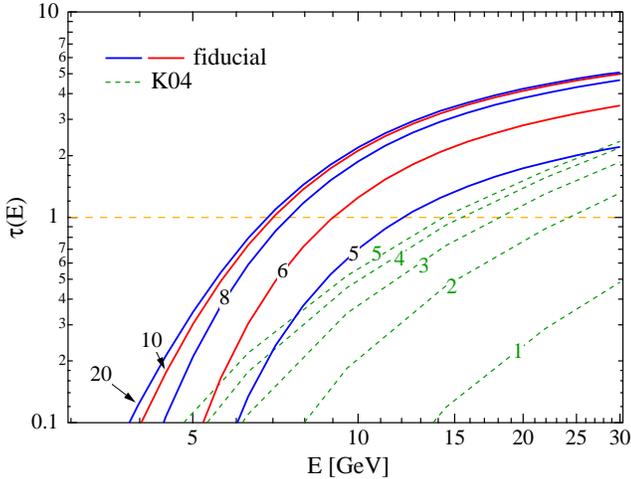,width=0.50\textwidth}
\caption{
Integrated $\gamma\gamma$ optical depth $\tau(E)$ vs. observed gamma-ray energy $E$
for source redshifts $z$ as labelled,
for the fiducial model (solid) and K04's high stellar UV model (dashed).
}
\label{fig:taudet}
\end{figure}

This can be seen more explicitly in Fig.\ref{fig:taudet} where we show
the integrated $\gamma\gamma$ optical depths for different source redshifts $z$,
\begin{eqnarray}
\tau (z,  E) = \int_{z_{\min}}^{z} dz' {dl \over dz'} \int_{\epsilon_{th}}^{\infty} d\epsilon \ n (\epsilon, z')\nonumber \\
                                 \times {1 \over 2} \int_{-1}^{1} d\mu (1 - \mu) \sigma (E(1+z'), \epsilon, \mu),
\end{eqnarray}
where $dl/dz'=l_H(z')/(1+z')$ and $E$ is the observed gamma-ray energy at $z=0$.
As mentioned above, the lower limit of $z$-integration that can be taken in our model is $z_{\min}=4$;
for additional absorption from the range $z=0-4$,
we can only consult other models at the moment
(e.g. Kneiske et al. 2004, hereafter K04; Stecker et al. 2006; Razzaque et al. 2009; Gilmore et al. 2009).
Overlayed here for comparison is K04's ``high stellar UV model'',
which gives their best description of QSO proximity effect measurements at $z \sim 2-4$.

As we could infer from Fig.\ref{fig:tauloc},
our model predicts appreciable opacity at observed energies $E \lesssim 12$ GeV for sources at $z \gtrsim 5$,
with notable differences out to $z \sim 8$.
However, the relative effects of further absorption from $z \gtrsim 8$ may be practically indiscernible.
Nevertheless, the spectral attenuation feature itself should be observable
in high-$z$ gamma-ray sources by current or future gamma-ray facilities,
and possibly distinguishable in the range $z \sim 5-8$ for sufficiently bright objects (\S \ref{sec:detect}).
Owing to the drop in $\gamma\gamma$ opacity at $E_{\rm rest} < 18$ GeV (Fig.\ref{fig:tauloc}),
the differences in absorption in this $z$ range are caused in-situ by the evolution of UV IRFs
just below the Lyman edge energy, including the crucial Ly$\alpha$ and Lyman-Werner bands.
We also recall that in this model,
Pop III stars continued to be significant contributors to the UV IRF down to $z \sim 7$,
where they are comparable with Pop II stars for ionizing photons.
Measurements of these effects would thus provide
an important check of current models of cosmic reionization in its latter stages,
as well as a unique and invaluable probe of evolving UV IRFs in the sub-Lyman edge regime
during the era of early star formation (\S \ref{sec:detect}).

\begin{figure}
\centering
\epsfig{file=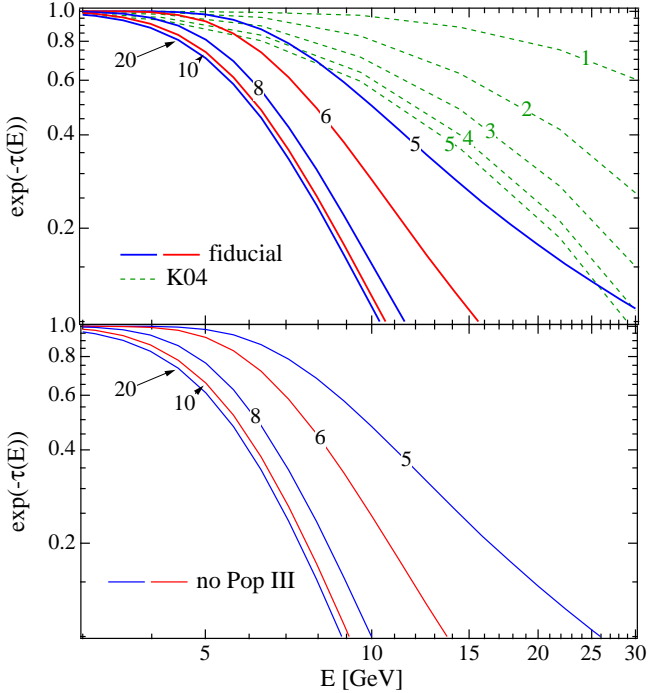,width=0.50\textwidth}
\caption{
Spectral attenuation factor $\exp[ -\tau(E)]$ vs. observed gamma-ray energy $E$
for source redshifts $z$ as labelled.
Upper panel: fiducial model (solid) and K04's high stellar UV model (dashed).
Lower panel: model without Pop III stars.
}
\label{fig:expdet_com}
\end{figure}

\begin{figure}
\centering
\epsfig{file=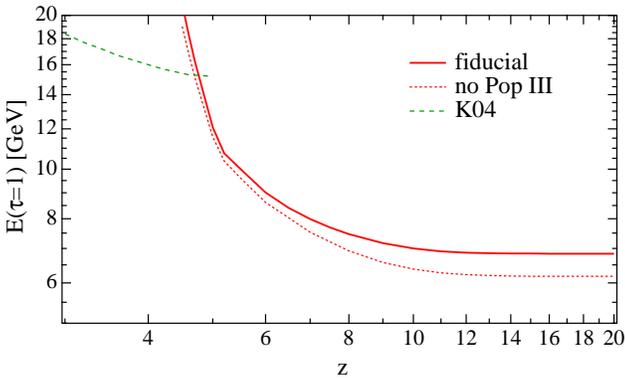,width=0.50\textwidth}
\caption{
Observed gamma-ray energy $E$ where $\tau=1$ vs. redshift $z$,
for the fiducial model (solid), model without Pop III stars (dotted)
and K04's high stellar UV model (dashed).
}
\label{fig:Etau1_com}
\end{figure}

In Figs. \ref{fig:expdet_com} and \ref{fig:Etau1_com}, respectively,
we plot the spectral attenuation factor $\exp [-\tau(E)]$
and the observed energy $E(\tau=1)$ where the optical depth is unity.
Here the fiducial results are compared with those of an alternative model
\footnote{The relevant parameters are $\varepsilon_{\rm *,II}=0.1$ and $f_{\rm esc,II}=0.0928$.
The model gives $\tau_e=0.06$, marginally consistent with the 5th year {\it WMAP} constraints.}
that does not include Pop III stars, in which reionization is driven only by Pop II stars
and occurs relatively late at $z \sim 6$ (similar to the late reionization model of Gallerani et al. 2008).
The fact that Pop II stars are less efficient sources of ionizing photons compared to Pop III stars
mandates a larger SFR, more intense IRF for $\epsilon < \epsilon_{\rm LE}$
and hence larger $\gamma\gamma$ opacity.
However, since the SFR at $z \lesssim 6$ is observationally constrained,
notable differences appear only at $z \gtrsim 8$, which should be challenging to distinguish in practice.
Thus gamma-ray absorption may not be a sensitive probe of the reionization history itself.
We have also investigated various other models,
e.g. those with more realistic prescriptions for radiative feedback
that fit the current high-$z$ observations nearly equally well,
and found that they generally do not lead to large differences.
Conversely, being constrained by existing data,
our predictions may be considered reasonably robust,
at least within the framework of our model.
Nevertheless, we caution that relaxing some of the present assumptions,
e.g. regarding the stellar IMF or the QSO contribution,
may yet allow a wider range of possibilities.
Note that
although some other recent models (e.g. Razzaque et al. 2009, Gilmore et al. 2009)
predict somewhat less absorption at $z \sim 5-6$,
they are not directly comparable with ours as their focus is on the $z<6$ universe
(e.g. Gilmore et al. 2009 do not attempt to fit the Ly$\alpha$ effective optical depths at $z \gtrsim 5.5$ as we do).

\section{Discussion}
\label{sec:detect}
The fact that $\gamma\gamma$ absorption is sensitive to photons with energies below the Lyman limit
rather than the ionizing radiation (\S \ref{sec:model}) actually points to a unique probe
of the cosmic reionization epoch that complements measurements
of QSO Gunn-Peterson troughs or CMB polarization anisotropies,
which probe the neutral and ionized components of the IGM, respectively.
On the one hand, observationally deducing the global UV emissivity and hence the cosmic star formation rate
from the latter two is problematic due to uncertainties in the inhomogeneity of the IGM (clumping factor)
and the escape fraction of ionizing photons from the host galaxies (Madau et al. 1999, Wyithe et al. 2009).
On the other, the direct census of the high-$z$ UV luminosity density from deep, near-IR surveys
are affected by the uncertain integrated contribution of faint galaxies below the telescope detection limit (e.g. Bouwens et al. 2007).
Observing $\gamma\gamma$ absorption in high-$z$ sources may allow more robust measurements
of the evolution of the cosmic UV emissivity, and in combination with other data,
possibly the determination of the escape fraction and/or the IGM clumping factor as well.
These inferences are general and independent of any particular model for reionization,
but will be investigated in more quantitative detail in the near future.
Likewise, the implications for constraining Ly$\alpha$ or $\rm H_2$-dissociating radiation
from $\gamma\gamma$ absorption will be discussed in future work. 

We now briefly address whether the effects discussed above are observable
in real sources with current or future gamma-ray instruments.
For blazars, the most prominent and numerous extragalactic sources of GeV gamma-rays,
the highest redshift confirmed so far
is $z \sim 3$ (Hartman et al. 1999; Abdo et al. 2009b).
However, objects similar to the most powerful known blazars such as 3C454.3
with apparent luminosities $L \sim 10^{49} {\rm erg\ s^{-1}}$
should be detectable by {\it Fermi} out to $z \sim 8-10$ if they exist at such redshifts (e.g. Romani et al. 2004).
According to the latest blazar evolution models (Inoue \& Totani 2009),
it may be plausible for {\it Fermi} to detect some blazars above $z \sim 6$ during its survey period,
for which deep, pointed observations may indeed reveal the IRF absorption features described above.

GRBs are also promising
as they are known to occur at $z>6$ (Kawai et al. 2005; Greiner et al. 2009),
at least up to $z \sim 8.2$ (Tanvir et al. 2009; Salvaterra et al. 2009),
and perhaps out to the very first epochs of star formation in the universe
(e.g. Bromm \& Loeb 2006; Salvaterra et al. 2008).
Although the spectral properties of GRBs in the GeV domain are still rather uncertain,
previous detections by CGRO/EGRET (Hurley et al. 1994)
and the recent detection of GRB 080916C at $z = 4.35$ by {\it Fermi} (Abdo et al. 2009a)
\footnote{For $z=4.35$ and $E=13.2$ GeV, our fiducial model gives $\tau_{\gamma\gamma} \simeq 0.4$,
consistent with the actual detection of a photon at this energy from GRB 080916C.}
demonstrate that at least some GRBs have luminous emission extending to $>$10 GeV,
which can also be expected theoretically (e.g. Zhang \& Meszaros 2001; Asano et al. 2009).
A burst similar to GRB 080916C may still be detectable at several GeV by {\it Fermi}/LAT at $z \lesssim 7$,
and even out to higher $z$ if the spectrum was somewhat harder.
Even better for this purpose
would be proposed ground-based telescopes with much larger effective area and multi-GeV energy threshold,
such as the Cherenkov Telescope Array (CTA)\footnote{http://www.cta-observatory.org},
Advanced Gamma-ray Imaging System (AGIS)\footnote{http://www.agis-observatory.org}
or the 5@5 array (Aharonian et al. 2001).
Together with measurements of Ly$\alpha$ damping wings (McQuinn et al. 2009)
and possibly radio dispersion (Ioka 2003; Inoue 2004),
future, broadband observations of very high-$z$ GRBs
should open new windows onto the cosmic reionization epoch.

Even if IRF-induced spectral features are detected,
a generic problem for $\gamma\gamma$ absorption studies
is distinguishing them from spectral cutoffs intrinsic to the source.
In this regard, spectral variability should offer an important clue.
Both blazars and GRBs are highly variable gamma-ray emitters,
and in general, changes in physical conditions of the source
that cause variations in flux should also be accompanied
by variations of the intrinsic cutoff energy,
whether it is due to injection of freshly accelerated particles,
changes in the magnetic fields, internal radiation fields, bulk flow velocity, etc.
In contrast, cutoffs of IRF origin should be stable in time
and independent of the variability state of each object.
Acquisition of time-resolved spectra
should thus allow the deconvolution of the two effects.
Another indication should come from statistical studies of a sufficient sample of measurements.
IRF-related cutoffs should occur at similar energies for sources at similar $z$,
and also exhibit a systematic evolution toward lower energies for higher $z$,
whereas there is no strong reason to expect such trends for intrinsic cutoffs.
Both the above strategies motivate the construction of future, high-sensitivity multi-GeV facilities such as CTA, AGIS and 5@5,
which should be powerful tools to probe the evolution of UV IRFs in the cosmic reionization era
through $\gamma\gamma$ absorption in very high-$z$ sources.

\section*{Acknowledgments}
We thank F. Aharonian, P. Coppi,
Y. Inoue, N. Kawai, F. Miniati, N. Omodei, J. Rhoads, M. Teshima and T. Totani for valuable discussions,
T. Kneiske for making her models available, and the anonymous referee for very helpful and constructive comments.
S. I. is supported by Grants-in-Aid for Scientific Research Nos. 19047004 and 19540283
and for the Global COE Program
"The Next Generation of Physics, Spun from Universality and Emergence"
from the Ministry of E.C.S.S.T. (MEXT) of Japan.


\end{document}